\documentclass[
reprint,     
superscriptaddress,
showkeys,
amsmath,amssymb,
aps,
pra,
floatfix,
]{revtex4-2}

\usepackage{natbib}
\usepackage{graphicx}
\usepackage{dcolumn}
\usepackage{bm}
\usepackage{svg}
\usepackage[hidelinks]{hyperref}
\usepackage[mathlines]{lineno}
\usepackage{multirow}
\usepackage{subcaption}

\usepackage{xcolor}
\usepackage{comment}
\usepackage{placeins}

\newcommand{\blue}[1]{\textcolor{blue}{#1}}
\newcommand{\period}{\,\mathrm{.}}
\newcommand{\comma}{\,\mathrm{,}}

\begin{document}

\title{Quantum Noise Reduction in the Space-based Gravitational Wave Antenna DECIGO Using Optical Springs and Homodyne Detection scheme}

\author{Kenji Tsuji}
\affiliation{Department of Physics, Nagoya University, Furo-cho, Chikusa-ku, Nagoya, Aichi 464-8602, Japan}

\author{Tomohiro Ishikawa}
\affiliation{Department of Physics, Nagoya University, Furo-cho, Chikusa-ku, Nagoya, Aichi 464-8602, Japan}

\author{Kentaro Komori}
\affiliation{Research Center for the Early Universe (RESCEU), School of Science, University of Tokyo, Tokyo 113-0033, Japan}
\affiliation{Department of Physics, University of Tokyo, Bunkyo, Tokyo 113-0033, Japan}

\author{Yutaro Enomoto}
\affiliation{Institute of Space and Astronautical Science, Japan Aerospace Exploration Agency, Sagamihara, Kanagawa 252-5210, Japan}

\author{Yuta Michimura}
\affiliation{Research Center for the Early Universe (RESCEU), School of Science, University of Tokyo, Tokyo 113-0033, Japan}
\affiliation{LIGO Laboratory, California Institute of Technology, Pasadena, California 91125, USA}

\author{Kurumi Umemura}
\affiliation{Department of Physics, Nagoya University, Furo-cho, Chikusa-ku, Nagoya, Aichi 464-8602, Japan}

\author{Shoki Iwaguchi}
\affiliation{Department of Physics, Nagoya University, Furo-cho, Chikusa-ku, Nagoya, Aichi 464-8602, Japan}

\author{Keiko Kokeyama}
\affiliation{Department of Physics, Nagoya University, Furo-cho, Chikusa-ku, Nagoya, Aichi 464-8602, Japan}
\affiliation{Cardiff University, Main Building, Park Place, Cardiff CF10 3AT, Wales, United Kingdom}
\affiliation{The Kobayashi-Maskawa Institute for the Origin of Particles and the Universe, Nagoya University, Nagoya, Aichi 464-8602, Japan}

\author{Seiji Kawamura}
\affiliation{Department of Physics, Nagoya University, Furo-cho, Chikusa-ku, Nagoya, Aichi 464-8602, Japan}


\date{\today}

\begin{abstract}
The DECi-hertz Interferometer Gravitational-wave Observatory (DECIGO) is a planned space-based, next-generation gravitational wave detector aimed at observing primordial gravitational waves originating form cosmic inflation. This work focuses on reducing the quantum noise, in the instrument's observation band of $0.1$ to $10$\,Hz, by employing optical springs and a homodyne detection scheme. Although detuning 1000\,km long arm cavities was previously considered ineffective due to quantum state degradation from diffraction losses, we revisit this problem by formulating a new, rigorous model for quantum state of light by accounting for the vacuum state mixing as a result of diffraction losses. This work shows that high sensitivities can be achieved by employing optimal configurations of optical springs and homodyne detection schemes even with diffraction losses. These improvements alone are still not sufficient to achieve sensitivities to detect primordial gravitational waves as other technical noises limit further improvement.
\end{abstract}
\maketitle

\clearpage
\section{Introduction}
Current long-baseline, terrestrial gravitational wave detectors of the LIGO-Virgo-KAGRA collaboration have detected multiple compact binary mergers around 100\,Hz. The first detection in 2015 \cite{PhysRevLett.116.061102}, and the detections from subsequent observation runs \cite{PhysRevX.13.041039, 10.1093/ptep/ptac073} have revolutionized astrophysics and helped improve population models and multi-messenger astronomy. Sensitivity towards gravitational waves in these frequencies is expected to improve by an order of magnitude with the planned next-generation detectors, such as the Einstein Telescope \cite{punturo2010einstein} and the Cosmic Explorer \cite{abbott2017exploring}. Even with the next-generation of detectors, the sensitivity at low frequencies around 1\,Hz is not expected to improve, and it is fundamentally limited by terrestrial seismic noise and test-mass suspension thermal noise.

DECIGO is a Japanese space-based gravitational wave antenna that is proposed to have optimal sensitivity for the frequency band $0.1$ to $10$\,Hz. This frequency band is particularly important as it bridges the gap between future space-based detectors operating in the mHz band, such as  the Laser Interferometer Space Antenna (LISA) \cite{AmaroSeoane2012,amaroseoane2017laserinterferometerspaceantenna}, and ground-based gravitational-wave detectors \cite{punturo2010einstein,abbott2017exploring}. Current design of DECIGO employs three drag-free spacecrafts that form optical cavities with a base-line of about 1000\,km \cite{PhysRevLett.87.221103,10.1093/ptep/ptab019,galaxies12020013}. A space-based gravitational wave observatory is free from the seismic noise, that is unavoidable by terrestrial interferometers and this advantage also extends to not needing to employ suspensions for vibration isolation of the test-masses, eliminating the suspension thermal noise contribution to the instrument noise. The main scientific goal of DECIGO is to detect primordial gravitational waves as a result of cosmic inflation \cite{MAGGIORE2000283, PhysRevD.37.2078, PhysRevD.42.453}, as their direct detection will yield key insights into early universe that is only possible to probe via gravitational waves.

The recent, \textit{Planck} cosmic microwave background measurements and others have set a new upper limit for allowed gravitational wave energy density, ${\Omega}_{\mathrm{GW}}$, by lowering it from $2{\times}10^{-15}$ to $1{\times}10^{-16}$ \cite{akrami2020planck}. As such, DECIGO's original design sensitivity is no longer sufficient to detect primordial gravitational waves, further emphasizing the need for considering new instrument design choices that can achieve better sensitivities.

The primary limiting noise source in DECIGO's observation band is quantum noise. Thus, a significant focus is placed on improving the instrument's quantum noise performance. In general, for cavities with very long baselines such as DECIGO, it has been assumed that diffraction losses prevent the preservation of correlated quantum states of light, reducing the effectiveness of techniques such as optical spring (via cavity detuning) and homodyne detection, which are commonly used in ground-based detectors. To address this issue, quantum locking schemes with short additional cavities have been proposed and analyzed \cite{YAMADA2020126626, YAMADA2021127365, galaxies11060111}. These studies demonstrate a significant improvement in quantum noise limited sensitivities by employing detuned, lossless sub-cavities along with the use of homodyne detection schemes.

A recent study has developed a framework to systematically model the impact of diffraction losses on the quantum states of light \cite{umemura2025introductionvacuumfieldscavity}. This enables a more consistent treatment than previous analyses, where diffraction losses were typically modeled only phenomenologically. This framework allows us to model the vacuum-state mixing induced by diffraction losses and evaluate the performance of quantum techniques such as optical springs and homodyne detection applied directly to the main long-baseline cavities of DECIGO. Using this framework, we evaluate the detector sensitivity in the observation band with detuning and homodyne detection, thereby clarifying how diffraction losses modify the performance predicted in earlier idealized treatments.

In section {\ref{sec:quantum_noise}}, we establish the treatment of quantum fluctuations used in this paper, following that influence of diffraction losses in a Fabry-Perot cavity, optical springs and homodyne detection schemes are also discussed. Section {\ref{sec:simulation}} presents the simulation method used, and search for optimal parameters for potential DECIGO design. This section also explains the constraints set by acceleration noise, which is considered a practical noise source, in search for optimal instrument design parameters. Following that, section {\ref{sec:result}} presents the optimized sensitivity for primordial gravitational waves using signal-to-noise ratio (SNR) as a metric. Finally, section {\ref{sec:summary}} concludes, with a future direction for this work.

\begin{figure*}[t]
  \centering
  \includegraphics[width=\textwidth]{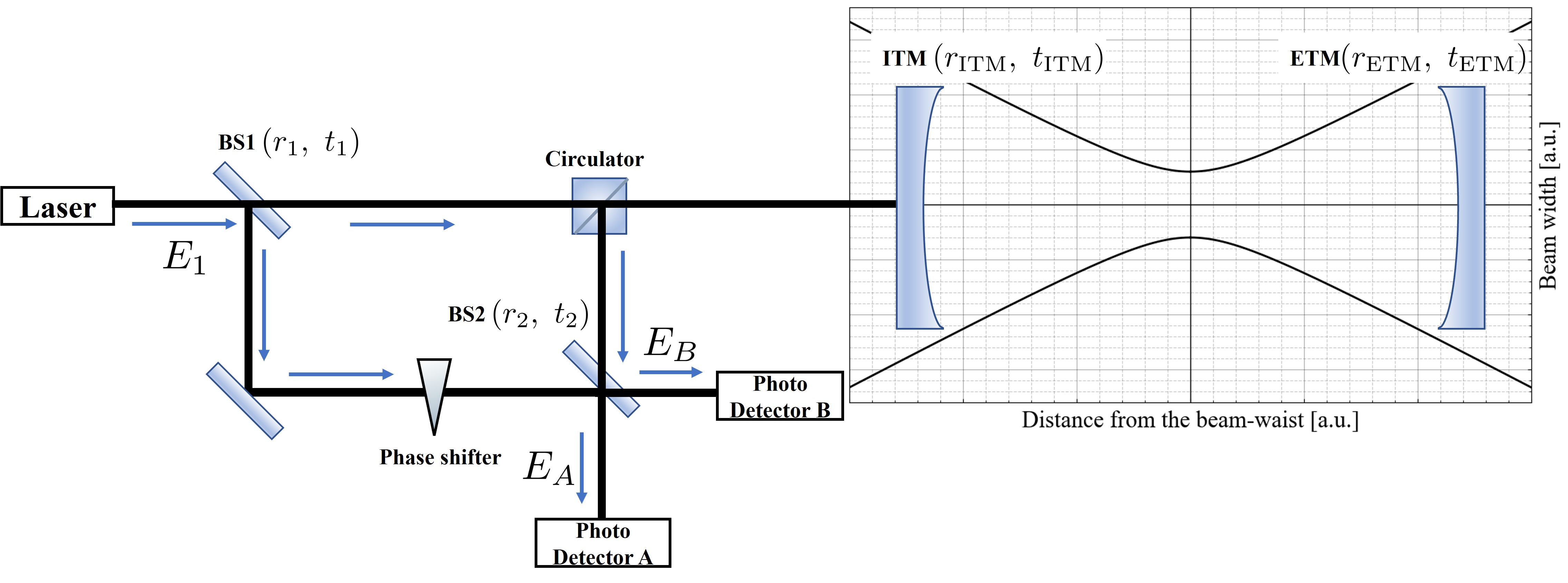}
  \caption{Optical design of homodyne detection scheme considering diffraction losses. BS1 is a beam splitter used to derive the Local Oscillator (LO) beam from the incident field $E_1$, with its parameters defined by the amplitude reflectivity $r_1$ and amplitude transmissivity $t_1$. The phase of the LO beam can be arbitrarily adjusted using a phase shifter. BS2 is a beam splitter with reflectivity $r_2$ and transmissivity $t_2$, used to interfere the light reflected from the arm cavity with the LO beam. Since the arm cavity is assumed to be long, as in DECIGO, diffraction losses become significant. The laser beam inside the cavity is configured to have its waist at the cavity's center, and the curvatures of the Input Test Mass (ITM) and End Test Mass (ETM) are assumed to be perfectly mode-matched to the wavefront of the Gaussian beam. The reflectivities and transmissivities of these two test masses are denoted as $r_{\mathrm{\scalebox{0.5}{ITM}}},t_{\mathrm{\scalebox{0.5}{ITM}}},r_{\mathrm{\scalebox{0.5}{ETM}}},t_{\mathrm{\scalebox{0.5}{ETM}}}$, respectively.
  }
  \label{fig:Path1}
\end{figure*}

\section{Quantum Noise in a Fabry–Perot Cavity with Diffraction Losses}\label{sec:quantum_noise}

In this section, we introduce a treatment of quantum fluctuations of light and describe how they manifest as quantum sensing noise that limit the sensitivity of DECIGO.

\begin{figure*}[t]
  \centering
  \includegraphics[width=\textwidth]{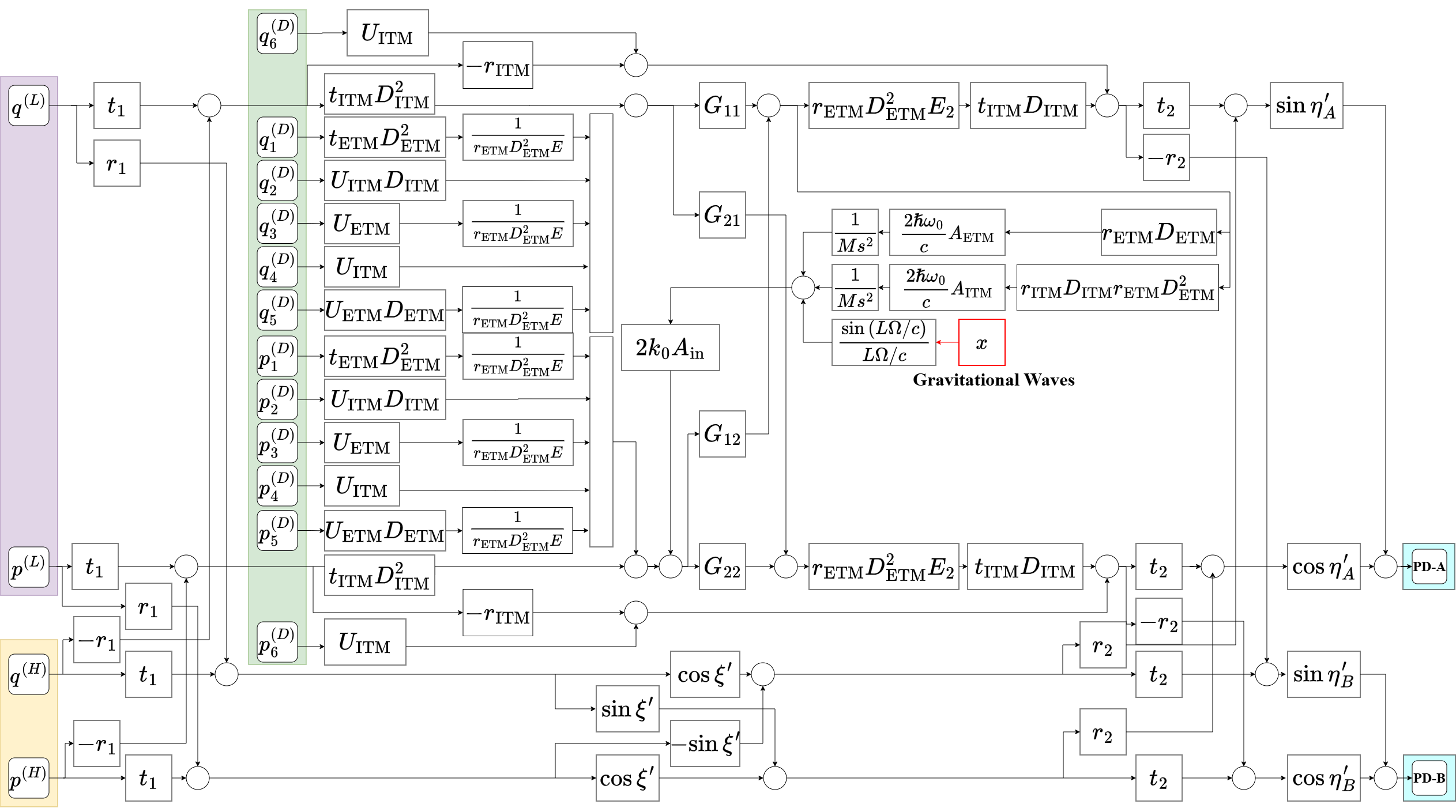}
  \caption{Block diagram for calculating the quantum noise of the detector, considering diffraction losses in the arm cavity. All input ports, except those associated with gravitational waves, represent quantum fluctuations entering the system. The two ports, $q^{\mathrm{(L)}}$ and $p^{\mathrm{(L)}}$ (shown in the purple region), correspond to the intrinsic quantum fluctuations of the laser beam, with the upper one representing amplitude quantum fluctuations and the lower one representing phase quantum fluctuations. The two ports, $q^{\mathrm{(H)}}$ and $p^{\mathrm{(H)}}$ (shown in the yellow region), correspond to vacuum quantum fluctuations entering through the open port of BS1, used to extract the LO beam for homodyne detection. The twelve input ports, $q^{\mathrm{(D)}}_i$ and $p^{\mathrm{(D)}}_i$(shown in the green region), represent vacuum quantum fluctuations entering the system due to diffraction losses \cite{umemura2025introductionvacuumfieldscavity}. The effect of gravitational waves enters the system as mirror displacement (shown in the red region). The two output ports, PD-A and PD-B (shown in the cyan region), correspond to the photodetectors.}
  \label{fig:Block1}
\end{figure*}

The optical setup for which the quantum noise limited sensitivity is calculated is shown in the Fig\,{\ref{fig:Path1}}. An input laser beam is split by a beam splitter (BS1), where, one part of light is coupled into a long Fabry-Perot cavity and the other part is used as a local oscillator (LO) for the homodyne detection scheme. The homodyne detection is implemented by interfering the reflection from the long cavity and the LO beam using another beam splitter (BS2). The light from the two output ports of BS2 are captured using two photodetectors (PD). The quantum noise coupling paths for this optical configuration are described in Fig\,{\ref{fig:Block1}}. Each input port corresponds to a point in the optical configuration where quantum fluctuations enter the system. These fluctuations propagate the graph, where blocks represent various optical components. The cumulative transfer function from each input port to the output port (the PDs) can be calculated by multiplying all the transfer functions of the optical components (blocks) in the traversed path.

\subsection{Quantum Fluctuation}\label{subsec:QF}
In this paper, we utilize the mathematical framework presented in {\cite{schleich2011quantum}} to describe quantum fluctuations. The commutation relation between the annihilation and creation operators $a$, $a^{\dagger}$ satisfy the follow relation -

\begin{equation}\label{eq:commutator}
  [a,a^{\dagger}] = 1\period
\end{equation}

The quantum fluctuation in amplitude and phase quadratures, q and p respectively, are defines as -

\begin{equation}
    q = \frac{1}{2}(a+a^{\dagger}),\space p=\frac{1}{2i}(a-a^{\dagger})\period
\end{equation}

The amplitude and phase quadrature quantum fluctuations from the laser source are given by $q^{\mathrm{(L)}}$ and $p^{\mathrm{(L)}}$ respectively.

\subsection{Diffraction Loss}\label{subsec:DL}
In case of a long Fabry-Perot cavity, such as the default design DECIGO, diffraction loss cannot be neglected. To model diffraction losses in a cavity, the notation from {\cite{umemura2025introductionvacuumfieldscavity}} is adopted. A parameterized factor $D$ is used to characterized diffraction losses, and is given by \cite{galaxies9010009} -

\begin{align}
    D^{2} &= 1-exp{\left( -\frac{2\pi}{L\lambda}R^{2}\right)}\\
    U^{2} &= 1 - D^{2}\comma
\end{align}
\noindent where $R$ is the mirror radius of input and end test masses, $L$ is the cavity length and $\lambda$ is the wavelength of the laser light. For the default design of DECIGO, $R=0.5$\,m, $L=1000$\,km and $\lambda=515$\,nm. The factor $D$ can be interpreted as the optical loss as a result of finite-size of the mirrors, as well as the loss that occurs when projecting the reflected optical mode of the cavity onto the fundamental Hermite-Gaussian mode, $\mathrm{TEM_{00}}$.
And, $U$ represents a parameterized factor that characterizes the optical mode preserved after accounting for diffraction losses.

The first type of loss arises from the finite aperture of the mirrors and is evaluated by integrating the spatial mode that is resonant in the cavity up to the mirror radius. This results in the reflected optical mode being a radially truncated Gaussian mode instead of being a pure Gaussian mode. In this study we assume an ideal cavity with only the $\mathrm{TEM_{00}}$ mode resonant, resulting in higher-order modes that are generated by this truncation to be naturally suppressed. The second type of loss, is a coupling mismatch of the input Gaussian mode to the truncated Gaussian mode that is circulating in the cavity. We consider $D$ as the factor that corresponds to the diffraction losses for the case where input mode parameters are optimal and maximize the laser power being delivered to the test masses, as described in {\cite{galaxies9010014}}. In the context of this optimization, it is assumed that the cavity geometry is fully mode-matched to the input Gaussian mode, with the mirror curvatures chosen accordingly.

We employ a rigorous treatment of quantum noise, where the quantum fluctuations associated with laser light lost due to diffraction are replaced with vacuum fluctuations, that co-propagate with the remaining mode. This results in the total quantum fluctuations in the system being preserved. The amplitude and phase quadratures of the vacuum fluctuations being mixed-in are represented by $q^{\mathrm{D}}_\mathrm{i}$ and $p^{\mathrm{D}}_\mathrm{i}$ respectively. Where $\mathrm{i}$ corresponds to the independent ports in Fig\,\ref{fig:Block1} (shown in the green region) where vacuum fluctuations are introduced into the system.

\subsection{Optical spring}\label{subsec:OS}
The circulating optical power in a Fabry-Perot cavity exerts a force on the test masses known as radiation pressure. Hence, to maintain the position of the test masses, a external balancing force is applied in feedback. When the cavity length is maintained at an offset greater than the resonance length, any outward displacement of test masses results in the intra-cavity power decreasing. This corresponds to a decrease of radiation pressure, which in-turn restores the cavity length to the original off-resonant length. This restoring behavior is known as optical spring. Conversely, when the cavity length in maintained at an offset less than the resonance length, any outward displacement of test masses increases the intra-cavity power, thus increasing the radiation pressure, and restoring the cavity length back to the original off-resonant length. This configuration of radiation pressure mediated restoring force is known as an optical anti-spring. In this work, we treat both of these configurations as same but with opposite signs of detuning angle, that characterizes the offset from resonant length of a cavity.

When operating a cavity with any optical spring configuration, amplitude and phase quadrature fluctuations are coupled through the cross terms of the following coupling matrix, $G$ -

\begin{align}
  G\left(\Omega\right) &= {\frac{1}{2}}
  \begin{bmatrix}
     g^{\ast}\left(\Omega\right)+g\left(-{\Omega}\right) & i\left\{g^{\ast}\left(\Omega\right)-g\left(-{\Omega}\right)\right\} \\
     -i\left\{g^{\ast}\left(\Omega\right)-g\left(-{\Omega}\right)\right\} & g^{\ast}\left(\Omega\right)+g\left(-{\Omega}\right)
  \end{bmatrix}\\
  g\left(\Omega\right) &= \left\{1-r_{\mathrm{\scalebox{0.5}{ITM}}}D_{\mathrm{\scalebox{0.5}{ITM}}}^2r_{\mathrm{\scalebox{0.5}{ETM}}}D_{\mathrm{\scalebox{0.5}{ETM}}}^2e^{-i\left({\phi}+{\frac{2L}{c}}{\Omega}\right)}\right\}^{-1}\period
\end{align}

\noindent Here, $r_{\mathrm{\scalebox{0.5}{ITM}}}$ and $r_{\mathrm{\scalebox{0.5}{ETM}}}$ represent the amplitude reflectivities of ITM and ETM mirrors, respectively. $\phi$ is the detuning angle, ${\Omega}$ is the sideband frequency, and $c$ is the speed of light ($3{\times}10^8$ m/s). Furthermore, the amplitude quadrature quantum fluctuations, mediated by radiation pressure, couple to the test mass motion via the transfer function $2{\hbar}{\omega}_0A_{\mathrm{l}}/c$, where $A_{\mathrm{l}}$ represents the amplitude of the light incident on each test mass and the subscript $l$ used as an identifier. According to the equation of motion, the test mass's displacement in response to this force exhibits a frequency dependence given by $1/ms^2$, where $m$ is the mass of the ITM or ETM, taken as $100$\,kg in the default DECIGO design, and $s$ is the complex frequency variable in the Laplace domain (or equivalently $s = i\omega$ in the Fourier domain).

Any displacement of test masses is coupled to phase quadrature fluctuations via $2A_{\mathrm{l}}k_{0}$, where $k_{0}$ is the optical sprint constant. As a result, the phase and amplitude quadrature quantum fluctuations become intrinsically coupled inside a detuned cavity, in the presence of an optical spring.

\subsection{Homodyne Detection}\label{subsec:HD}
Homodyne detection scheme allows for selective readout of quantum noise along any quadrature in phase-amplitude phase space. The amplitude of the electric field from the laser source, $E_{\mathrm{l}}$, is defined as -

\begin{equation}
    E_{\mathrm{l}}=E_{0} e^{i{\omega}_0 t}\comma
\end{equation}

\noindent where $E_{0}$ is a constant and $\omega_{0}$ is the angular frequency of the laser light. The amplitude of the electric fields at the detectors PD-A and PD-B, represented by $E_{\mathrm{A}}$ and $E_{\mathrm{B}}$, are given by \cite{galaxies11060111} -

\begin{widetext}
    \begin{align}
        \label{eq:E_A}
        E_A &= \left(A_1+iA_2\right)E_1, \space \mathrm{where -}\\
        A_1 &= t_1t_2D_{\mathrm{\scalebox{0.5}{ITM}}}\left\{-r_{\mathrm{\scalebox{0.5}{ITM}}}+{\frac{t_{\mathrm{\scalebox{0.5}{ITM}}}^2D_{\mathrm{\scalebox{0.5}{ITM}}}^2r_{\mathrm{\scalebox{0.5}{ETM}}}D_{\mathrm{\scalebox{0.5}{ETM}}}^2\left({\cos{{\phi}^{\prime}}}-r_{\mathrm{\scalebox{0.5}{ITM}}}D_{\mathrm{\scalebox{0.5}{ITM}}}^2r_{\mathrm{\scalebox{0.5}{ETM}}}D_{\mathrm{\scalebox{0.5}{ETM}}}^2\right)}{\left(1-r_{\mathrm{\scalebox{0.5}{ITM}}}D_{\mathrm{\scalebox{0.5}{ITM}}}^2r_{\mathrm{\scalebox{0.5}{ETM}}}D_{\mathrm{\scalebox{0.5}{ETM}}}^2\right)^2+4r_{\mathrm{\scalebox{0.5}{ITM}}}D_{\mathrm{\scalebox{0.5}{ITM}}}^2r_{\mathrm{\scalebox{0.5}{ETM}}}D_{\mathrm{\scalebox{0.5}{ETM}}}^2\ {\sin^2{{\frac{{\phi}^{\prime}}{2}}}}}}\right\}+r_1r_2\ {\cos{\xi}}\\
        A_2 &= t_1t_2D_{\mathrm{\scalebox{0.5}{ITM}}}\left\{{\frac{t_{\mathrm{\scalebox{0.5}{ITM}}}^2D_{\mathrm{\scalebox{0.5}{ITM}}}^2r_{\mathrm{\scalebox{0.5}{ETM}}}D_{\mathrm{\scalebox{0.5}{ETM}}}^2\ {\sin{{\phi}^{\prime}}}}{\left(1-r_{\mathrm{\scalebox{0.5}{ITM}}}D_{\mathrm{\scalebox{0.5}{ITM}}}^2r_{\mathrm{\scalebox{0.5}{ETM}}}D_{\mathrm{\scalebox{0.5}{ETM}}}^2\right)^2+4r_{\mathrm{\scalebox{0.5}{ITM}}}D_{\mathrm{\scalebox{0.5}{ITM}}}^2r_{\mathrm{\scalebox{0.5}{ETM}}}D_{\mathrm{\scalebox{0.5}{ETM}}}^2\ {\sin^2{{\frac{{\phi}^{\prime}}{2}}}}}}\right\}+r_1r_2\ {\sin{\xi}}
    \end{align}
    \begin{align}
        \label{eq:E_B}
        E_B &= \left(B_1+iB_2\right)E_1, \space \mathrm{where -}\\
        B_1 &= -t_1r_2D_{\mathrm{\scalebox{0.5}{ITM}}}\left\{-r_{\mathrm{\scalebox{0.5}{ITM}}}+{\frac{t_{\mathrm{\scalebox{0.5}{ITM}}}^2D_{\mathrm{\scalebox{0.5}{ITM}}}^2r_{\mathrm{\scalebox{0.5}{ETM}}}D_{\mathrm{\scalebox{0.5}{ETM}}}^2\left({\cos{{\phi}^{\prime}}}-r_{\mathrm{\scalebox{0.5}{ITM}}}D_{\mathrm{\scalebox{0.5}{ITM}}}^2r_{\mathrm{\scalebox{0.5}{ETM}}}D_{\mathrm{\scalebox{0.5}{ETM}}}^2\right)}{\left(1-r_{\mathrm{\scalebox{0.5}{ITM}}}D_{\mathrm{\scalebox{0.5}{ITM}}}^2r_{\mathrm{\scalebox{0.5}{ETM}}}D_{\mathrm{\scalebox{0.5}{ETM}}}^2\right)^2+4r_{\mathrm{\scalebox{0.5}{ITM}}}D_{\mathrm{\scalebox{0.5}{ITM}}}^2r_{\mathrm{\scalebox{0.5}{ETM}}}D_{\mathrm{\scalebox{0.5}{ETM}}}^2\ {\sin^2{{\frac{{\phi}^{\prime}}{2}}}}}}\right\}+r_1t_2\ {\cos{\xi}}\\
        B_2 &= -t_1r_2D_{\mathrm{\scalebox{0.5}{ITM}}}\left\{{\frac{t_{\mathrm{\scalebox{0.5}{ITM}}}^2D_{\mathrm{\scalebox{0.5}{ITM}}}^2r_{\mathrm{\scalebox{0.5}{ETM}}}D_{\mathrm{\scalebox{0.5}{ETM}}}^2\ {\sin{{\phi}^{\prime}}}}{\left(1-r_{\mathrm{\scalebox{0.5}{ITM}}}D_{\mathrm{\scalebox{0.5}{ITM}}}^2r_{\mathrm{\scalebox{0.5}{ETM}}}D_{\mathrm{\scalebox{0.5}{ETM}}}^2\right)^2+4r_{\mathrm{\scalebox{0.5}{ITM}}}D_{\mathrm{\scalebox{0.5}{ITM}}}^2r_{\mathrm{\scalebox{0.5}{ETM}}}D_{\mathrm{\scalebox{0.5}{ETM}}}^2\ {\sin^2{{\frac{{\phi}^{\prime}}{2}}}}}}\right\}+r_1t_2\ {\sin{\xi}}
    \end{align}
\end{widetext}

\noindent here, $r_{1}$ and $t_{1}$ are the amplitude reflectivity and transmissivity of BS1, while $r_{2}$ and $t_{2}$ are those of the beam splitter, BS2, used in the homodyne detector for recombining the LO and the reflected beam fron the cavity. In this study, the optics in the homodyne detection system are considered to be ideal and lossless. Thus, for these optics, the following equation is satisfied -

\begin{equation}
    r^{2} + t^{2} = 1\period
\end{equation}

\noindent In addition,  ${\phi}^{\prime}$ represents the detuning angle from resonance and $\xi$ represents the relative phase shift of the LO light with respect to $E_1$, which is dues for choosing the readout quadrature of the homodyne detector. Unlike previous studies \cite{galaxies11060111}, we include the parameter $D$, which accounts for the loss in carrier light in the reflection of the cavity due to diffraction.

The phase of the carrier light just before BS2 is denoted by $\eta_{0}$, the relative angles $\eta_{A}$ and $\eta_{B}$ with respect to the axes onto which each quadrature of quantum fluctuation is projected is given by -

\begin{align}
  {\eta}_{A} &= {\frac{\pi}{2}}+{\arctan{\left(\frac{A_1}{A_2}\right)}}-{\eta}_0\\
  {\eta}_{B} &= {\frac{\pi}{2}}+{\arctan{\left(\frac{B_1}{B_2}\right)}}-{\eta}_0\period
\end{align}
These angels are determined by specific optical parameters, and since the carrier light is included in the light coming from the cavity, it is not possible to freely choose the angles.

In this optical setup, vacuum fluctuations associated with homodyne detection enter the system through the unused port of BS1. The amplitude and phase quadratures of these fluctuations are denoted by $q^{\mathrm{(H)}}$ and $p^{\mathrm{(H)}}$, respectively.

\section{Simulations and optimal parmeter search}\label{sec:simulation}
\subsection{Signal Processing}

We define the signal corresponding to unit gravitational waves, $\textbf{x}$, in the cavity, as seen by the photodetectors PD-A and PD-B as $V_{A}$ and $V_{B}$, respectively -

\begin{align}
  V_{A} &= {\alpha}{\textbf{x}}+{\sum_{j}}a_j{\textbf{n}}_{j}\\
  V_{B} &= {\beta}{\textbf{x}}+{\sum_{j}}b_j{\textbf{n}}_{j}\comma
\end{align}

\noindent where $\alpha$ and $\beta$ are the transfer function from the displacement caused by gravitational waves in the cavity length to the detectors PD-A and PD-B, respectively. Similarly, $\textbf{n}_{j}$ represents a unit quantum fluctuation that is influenced by the transfer functions $a_j$ or $b_j$ to reach PD-A or PD-B. $\textbf{n}_{j}$ can be one of the following $q^{\mathrm{(L)}}$, $q^{\mathrm{(D)}}_i$, $q^{\mathrm{(H)}}$, $p^{\mathrm{(L)}}$, $q^{\mathrm{(D)}}_i$, $q^{\mathrm{(H)}}$.

In \cite{YAMADA2020126626}, an optimization method for extracting signals is proposed. A new signal $V$ is proposed as a combination of two arbitrary complex signals $V_{1}$ and $V_{2}$ and a combination coefficient $\chi$. The combination relation is -

\begin{equation}
\label{eq:new_signal}
  \begin{split}
    V &= V_1+{\chi}V_2\period
  \end{split}
\end{equation}

\noindent here, only $V_{1}$ contains the unit gravitational wave signal, $\textbf{x}$, and $V_{2}$ contains the any noise contributions in the signal and not $\textbf{x}$. The signal is optimized by determining ${\chi}={\chi}^{\mathrm{(opt)}}$, which minimizes the absolute value of $V$ according to eq.\,\ref{eq:opt_signal} -

\begin{equation}
\label{eq:opt_signal}
  \begin{split}
    S_{\mathrm{qua}} &= \left|V_1\right|^2+\left|V_2\right|^2\left|{\chi}+{\frac{V_2^{\dag}V_1}{V_2^{\dag}V_2}}\right|^2-{\frac{|V_2^{\dag}V_1|^2}{|V_2|^2}}\\
    S_{\mathrm{qua}}^{\mathrm{(opt)}} &= \left|V_1\right|^2-{\frac{\left|V_2^{\dag}V_1\right|^2}{\left|V_2\right|^2}}\hspace{2mm}\left(\mathrm{where}\hspace{2mm}{\chi}=-{\frac{V_2^{\dag}V_1}{V_2^{\dag}V_2}}\right)\period
  \end{split}
\end{equation}
Note that $S_{\mathrm{qua}}$ denotes the absolute square of $V$. To ensure that $V_2$ does not contain $\textbf{x}$, $V_1$ and $V_2$ are defined as a combination of $V_A$ and $V_B$ as follows -

\begin{equation}
  \begin{split}
    V_1 &=  {\frac{1}{\alpha}}V_A = {\textbf{x}}+{\frac{1}{\alpha}}{\sum_{j}}a_j{\textbf{n}}_{j}\\
    V_2 &= {\frac{1}{\alpha}}\left({\beta}V_A-{\alpha}V_B\right)={\frac{\beta}{\alpha}}{\sum_{i}}a_j{\textbf{n}}_{j}-{\sum_{j}}b_j{\textbf{n}}_{j}\\
    &= {\sum_{j}}\left({\frac{\beta}{\alpha}}a_j-b_j\right){\textbf{n}}_{j}\period
  \end{split}
\end{equation}

\noindent As $\alpha$ represents the transfer function of gravitational wave signal, dividing $V_{1}$ and $V_{2}$ by $\alpha$ calibrates the entire signal with respect to the gravitational wave signal. To eliminate the dependence of $\textbf{x}$ in $V_{2}$ we multiply $V_{A}$ by $\beta$ and $V_{B}$ by $\alpha$ and then subtract them.

\subsection{Acceleration Noise}\label{subsec:acc_noise}
To optimize the arm length and cavity finesse as free parameters in DECIGO, with the goal to improve quantum noise performance, acceleration noise should also be considered. Acceleration noise is a practical noise source caused by white forces acting on the test masses. It exhibits a frequency dependency that follows a $1/f^{2}$ trend. In the absence of this noise source, shortening the arm length and compensating for the reduced GW signal by increasing the finesse would not pose a disadvantage, as shot noise could be reduced without limit by increasing the circulating power.

Thus, in calculating the detector sensitivity, we consider acceleration noise, which is inversely proportional to arm length $L$. In this study, we define the acceleration noise of DECIGO, $S_{\mathrm{acc}}$, using the requirement factor ${\varepsilon}$, which is a factor relative to the radiation pressure noise level ${\mathcal{N}}_{\mathrm{RP},0.1}$ at $0.1$\,Hz from the default design, given by -

\begin{equation}
  S_{\mathrm{acc}} = {\varepsilon}^2{\mathcal{N}}_{\mathrm{RP},0.1}^2\left({\frac{1000{\mathrm{\,km}}}{L}}\right)^2\left({\frac{0.1{\mathrm{\,Hz}}}{f}}\right)^4.
\end{equation}

The total power spectral density, $P$, of the total detected signal is given by -

\begin{equation}
  P = S_{\mathrm{qua}}^{\mathrm{(opt)}} + S_{\mathrm{acc}}\hspace{1mm}.
\end{equation}

If we set a negligibly small heuristic for the acceleration noise, at $5\%$ of the radiation pressure noise, this corresponds to $\varepsilon=10^{-1/2}$. In this study, we adopt this value of $\varepsilon$ as the representative value for acceleration noise for the default DECIGO design and use it as a reference to compare with optimized results.

\subsection{Simulation Condition}

\begin{table}[t]
  \centering
  \caption{Conditions of parameters for the optimization of the DECIGO's sensitivity.}
  \label{table:parameter condition}
  \begin{tabular}{lll}
    \hline
    \hline
    Meaning & Symbol  &  \multicolumn{1}{c}{Range/Value}\\
    \hline
    \hline
    \multirow{2}{*}{Amplitude Transmissivity} & $t_1^{(h)}$ & $0$ to $1$\\
    &$t_2^{(h)}$ & ${\frac{1}{\sqrt{2}}}$ (Fixed)\\[2mm]
    Detuning Angle & ${\phi}$ & $-{\pi}$ to $\pi$\ rad\\[2mm]
    Finesse & ${\mathcal{F}}$ & $1$ to $500$\\[2mm]
    Phase Shift & ${\xi}$  & $-{\frac{\pi}{2}}$ to ${\frac{\pi}{2}}$\ rad\\[2mm]
    Arm Length & $L$ & $100$ to $1000$\ km\\[2mm]
    Acceleration Noise Factor & $\varepsilon$ & $\left(10^{-\frac{1}{2}},\ 10^{-1},\ 10^{-\frac{3}{2}},\ 10^{-2}\right)$\\
    \hline
    Laser Power & $I$ & $100$\ W\\[2mm]
    Laser Wavelength & ${\lambda}$ & $515$\ nm\\[2mm]
    Mirror Mass & $M$ & $100$\ kg \\[2mm]
    \hline
    \hline
  \end{tabular}
\end{table}

\begin{table}[t]
  \centering
  \caption{Parameters used to calculate the SNR.}
  \label{table:Parameters used to calculate the SNR}
  \begin{tabular}{lll}
    \hline
    Meaning & Symbol  &  \multicolumn{1}{c}{Value}\\
    \hline
    Hubble Parameter & $H_0$ & $70{\mathrm{\,km{\cdot}sec^{-1}{\cdot}Mpc^{-1}}}$\\
    Time for Correlation & T & 3 years\\
    Frequency & $f$ & 0.1 to 1 Hz\\
    Correlation Function & ${\gamma}$ & 1\\
    Energy Density & ${\Omega}_{\mathrm{GW}}$ & $10^{-16}$\\
    Noise Power Spectral Densities &$P_1,P_2$ &\\
    \hline
  \end{tabular}
\end{table}

In this section, we describe the parameters and their ranges used in the simulation to evaluate the sensitivity of DECIGO. Table\,\ref{table:parameter condition} summarizes  the parameters and their corresponding ranges. The bottom three parameters are kept constant in all simulations. The top 5 parameters are varied to evaluate for sensitivity and find the optimal value. 

The five parameters are - amplitude transmissivities of the beam splitters, $t_{1}^{(h)}$ and $t_{2}^{(h)}$. The detuning angle, $\phi$, that parameterize the optical spring effect. Finesse of the DECIGO arm cavity, ${\mathcal{F}}$. The phase difference between the carrier light reflected from the cavity and the LO, $\xi$. 

The five parameters are: amplitude transmissivities of the beam splitters, $t_{1}^{(h)}$ and $t_{2}^{(h)}$; the detuning angle, $\phi$, that parameterizes the optical spring effect; finesse of the DECIGO arm cavity, ${\mathcal{F}}$; the phase difference between the carrier light reflected from the cavity and the LO, $\xi$.
As the detectors PD-A and PD-B are positioned on the opposite output ports of BS-2, the phase difference between the two detectors is $\xi$ and $\xi + \pi$ respectively. So, it is sufficient to simulate $\xi$ in the range between $0$ and $\pi$. Additionally, the acceleration noise factor, $\varepsilon$, is varied from a list of fixed values - $\left(10^{-1/2}, 10^{-1}, 10^{-3/2},10^{-2}\right)$, as shown in the table.

To reduce the computational load and efficiency optimizing in between multiple parameters simultaneously, each signal is calculated using algebraic solutions derived from the block diagram shown in Fig\,\ref{fig:Block1}. Next, we adopt Signal-to-Noise Ratio (SNR) as a measure to evaluate the sensitivity of DECIGO towards primordial gravitational waves. The SNR is given by \cite{PhysRevD.59.102001}:

\begin{equation}
\label{eq:SNR}
  \mathrm{SNR} = {\frac{3H_0^2}{10{\pi}^2}}{\sqrt{T}}\left[{\int_{0.1}^1}df{\frac{2{\gamma}(f)^2{\Omega}^2_{\mathrm{GW}}(f)}{f^6P_1(f)P_2(f)}}\right]^{1/2}\comma
\end{equation}
\noindent where $P_1$ and $P_2$  represent the power spectral densities of the noise, calculated following the methodology outlined in section \ref{subsec:acc_noise}. In this case, $P_1$ and $P_2$ are assumed to be equal.

The remaining parameters used in the SNR calculation are summarized in Table\,\ref{table:Parameters used to calculate the SNR}. In Eq\,\ref{eq:SNR}, $T$ represents the observation period, which is set to 3 years. ${\Omega}_{\mathrm{GW}}$ denotes the energy density of primordial gravitational waves and is fixed at $10^{-16}$, this would allow for this measurement to set new upper limits. Furthermore, the overlap reduction function $\gamma$ \cite{PhysRevD.59.102001} is assumed to be unity for the DECIGO configuration. The SNR calculation focuses specifically on the frequency range from $0.1$\,Hz to $1$\,Hz, the optimal sensitivity range of DECIGO. Here, the lower bound of $0.1$\,Hz is determined by the presence of confusion foreground noise, primarily originating from white dwarf binaries \cite{10.1111/j.1365-2966.2003.07176.x}.

\begin{figure*}[t]
  \centering
  \includegraphics[width=0.7\textwidth]{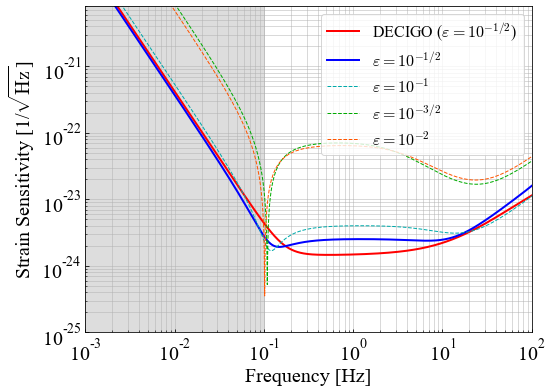}
  \caption{Optimized sensitivity curve for each acceleration noise level, along with the sensitivity in the case without an optical spring and homodyne detection. The two solid lines represent the difference in sensitivity for a fixed acceleration noise level, $\varepsilon=10^{-1/2}$, between configurations with and without homodyne detection and optical spring. The blue solid line corresponds to the case with both the optical spring and homodyne detection, while the red solid line corresponds to the case without them, which corresponds to the default DECIGO design. The dashed lines illustrate the sensitivity curves obtained by varying the value of $\varepsilon$ while employing homodyne detection and optical spring. Note that the gray-shaded region below 0.1 Hz represents the confusion noise, and this frequency region is excluded from the present analysis.}
  \label{fig:DECIGO_Defalut}
\end{figure*}

\begin{table*}[t]
  \centering
  \caption{Optimized Parameters.}
  \label{table:Optimized_Parameters}
  \begin{tabular}{c|lllllll||l}
    \hline
    \hline
    & $\varepsilon$ & $r_1$ & $r_2$ & ${\phi}$ (rad)& ${\mathcal{F}}$ & ${\xi}$ (rad)& $L$ (km) & SNR\\
    \hline
    \hline
    DECIGO (Default)&$10^{-{\frac{1}{2}}}$  & $-$ & $-$ & $-$  & $10.0$ & $-$  & $1000$ & $4.04$\\[2mm]
    \hline
              &$10^{-{\frac{1}{2}}}$ & $0.280$ & ${\frac{1}{\sqrt{2}}}$ & $0.135$ & $12.5$ & $-0.995$ & $900$ & $6.24$\\[2mm]
    Optimized &$10^{-1}$             & $0.111$ & ${\frac{1}{\sqrt{2}}}$ & $0.173$ & $11.2$ & $1.52$   & $675$ & $8.18$\\[2mm]
              &$10^{-{\frac{3}{2}}}$ & $0.385$ & ${\frac{1}{\sqrt{2}}}$ & $0.064$ & $17.5$ & $0.045$  & $225$ & $14.9$\\[2mm]
              &$10^{-2}$             & $0.381$ & ${\frac{1}{\sqrt{2}}}$ & $0.053$ & $17.5$ & $0.943$ & $225$ & $33.4$\\[2mm]
    \hline
    \hline
  \end{tabular}
\end{table*}

\begin{figure*}[!ht]
  \begin{minipage}{0.49\textwidth}
    \centering
    \includegraphics[width=\textwidth]{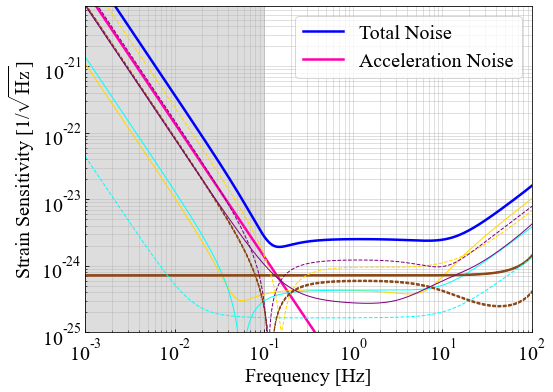}
    \subcaption{$\varepsilon=10^{-1/2}$}
    \label{fig:acc_10_to_negative_one_half}
  \end{minipage}
  \begin{minipage}{0.49\textwidth}
    \centering
    \includegraphics[width=\textwidth]{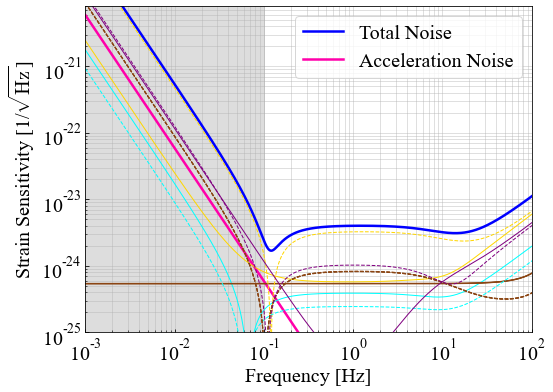}
    \subcaption{$\varepsilon=10^{-1}$}
    \label{fig:acc_10_to_negative_one}
  \end{minipage}\\
  \begin{minipage}{0.49\textwidth}
    \centering
    \includegraphics[width=\textwidth]{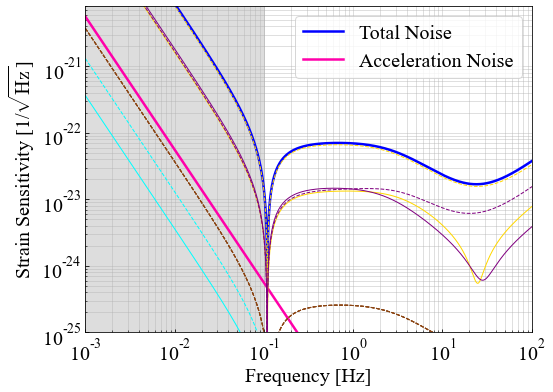}
    \subcaption{$\varepsilon=10^{-3/2}$}
    \label{fig:acc_10_to_negative_three_halves}
  \end{minipage}
  \begin{minipage}{0.49\textwidth}
    \centering
    \includegraphics[width=\textwidth]{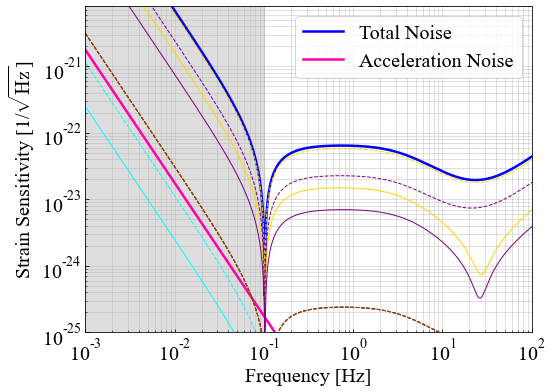}
    \subcaption{$\varepsilon=10^{-2}$}
    \label{fig:acc_10_to_negative_two}
    \end{minipage}
    \caption{Optimized sensitivity curves for each acceleration noise factor, $\varepsilon$, with contributions from various noise sources. In all subfigures, the total detector noise is represented by a thick blue curve, while the acceleration noise is shown as a thick magenta curve. The thin solid and dashed curves indicate the contributions from the quantum fluctuation defined in Figure \ref{fig:Block1}. The solod curves correspond to phase fluctuations $p$, while the dashed curves correspond to amplitude fluctuations $q$. The yellow curves correspond to the intrinsic quantum field of the laser light, namely $p^{(L)}$ and $q^{(L)}$. The purple curves represent the vacuum fields introduced through homodyne detection, corresponding to $p^{(H)}$ and $q^{(H)}$. The contributions from $p^{(D)}_1$ and $q^{(D)}_1$, which enter from behind the ETM, are negligibly small and lie outside the display range. The fields $p^{(D)}_{i}$ and $q^{(D)}_{i}$ ($i \in 2,3,4,5$) are vacuum fields associated with intracavity losses and are shown as dark brown curves. Specifically, $i=2$ corresponds to diffraction at the ITM; $i=3$, to HOM coupling at the ETM; $i=4$, to HOM coupling at the ITM; and $i=5$, to diffraction at the ETM. Since the coefficients of the transfer functions for these quantum fluctuations are very similar, all four components $p^{(D)}_{2}$, $p^{(D)}_{3}$, $p^{(D)}_{4}$, and $p^{(D)}_{5}$ (and similarly $q^{(D)}_{2}$, $q^{(D)}_{3}$, $q^{(D)}_{4}$, and $q^{(D)}_{5}$) visually overlap in Figures \ref{fig:acc_10_to_negative_one_half} and \ref{fig:acc_10_to_negative_one}. The cyan curves represent the vacuum fields entering from the detection port side of the ITM, corresponding to $p^{(D)}_6$ and $q^{(D)}_6$.}
  \label{fig:Optimized_sensitivity_curves}
\end{figure*}

\section{Results and Discussion}\label{sec:result}

In Fig\,\ref{fig:DECIGO_Defalut} the optimal sensitivity curves for various acceleration noise levels are shown, and the reference sensitivity plot of default DECIGO design is also shown. In the absence of an optical spring and a homodyne detection scheme, DECIGO's sensitivity is limited by radiation pressure noise, which follows a frequency trend of $1/f^{2}$, and shot noise, which is flat up to about $10$ Hz and increases proportionally to $f$ above that. Introducing an optical spring and homodyne detection allows for the quantum noise contribution to be suppressed at a specific frequency.

If optimizing SNR towards frequency dependent primordial gravitational waves is the goal, it is advantageous to form a narrow region of high sensitivity (a dip) at lower frequencies. However, if the dip appears below $0.1$\,Hz, the sensitivity cannot be improved due to the influence of confusion foreground noise. As a result, all optimized sensitivity curves exhibit a dip near $0.1$\,Hz, regardless of the acceleration noise level. While the exact shape of the dip varies as a function of the acceleration noise level, the general trend is that the dip becomes deeper and sharper as the acceleration noise decreases. A detailed explanation of these differences is provided later.

The optimized parameters are summarized in Table {\ref{table:Optimized_Parameters}}. When comparing the optimal sensitivities at acceleration noise factor $\varepsilon=10^{-1/2}$, with and without optical spring and homodyne readout scheme, the simulated sensitivity corresponds to an SNR of $6.24$ and $4.04$ respectively. This is an improvement factor of about $1.5$ times.

Next, we compare the optimized sensitivities at different levels of acceleration noise. From the perspective of just SNR, lower acceleration noise leads to higher, optimized SNRs. At $\varepsilon=10^{-2}$, the optimal SNR reaches $33.4$. This SNR is approximately 40\% of the value of 79.6 obtained in previous studies under idealized conditions without acceleration noise, where optimized quantum locking combined with optical springs and homodyne detection was considered \cite{galaxies11060111}. While the absolute SNR is lower, the present approach does not require additional large-scale optical components. In quantum locking schemes, sub-cavities with mirror masses comparable to the main test masses are typically required. Therefore the present method can be implemented within the existing DECIGO-like configuration, providing advantages in terms of design flexibility and practical feasibility for mission implementation. However, as shown in Fig\,\ref{fig:acc_10_to_negative_one} and Fig\,\ref{fig:acc_10_to_negative_three_halves}, when $\varepsilon$ is less than $10^{-1}$, the optimal sensitivity curve develops a sharp dip at $0.1$\,Hz. Although, this allows for strong probing of the primordial GWs at this frequency, it comes at the cost of worse sensitivity in the broadband, posing significant limitation is extracting information from other types of GWs. Given its broad science goals, therefore, these optical configurations are considered unsuitable for DECIGO.

Additionally, other practical noise sources, like the thermal noise caused by residual gas inside the satellite should be considered. Although the gas pressure varies depending on several factors, assuming a representative value of $10^{-9}$ Pa, as used in previous studies, the optimized sensitivity with $\varepsilon = 10^{-1}$ shows only marginal degradation. This suggests that searching for optical configurations at $\varepsilon$ values less than $10^{-1}$ is impractical due to the increasing impact of residual gas thermal noise. Therefore, in this study, we limit our search to acceleration noise factors satisfying ${\varepsilon}{\geq}10^{-1}$. Under this condition, the maximum achievable SNR using this method is $8.18$. Although this corresponds to approximately a twofold improvement compared with the configuration without optical springs and homodyne detection, it still remains insufficient for the robust detection of primordial gravitational waves. Since the optimized parameters and achievable SNR depend on both the amplitude and spectral shape of the assumed noise model, it would therefore be important to further investigate the contributions from additional realistic noise sources, particularly around the sensitivity dip region which gives the dominant contribution to the SNR.

Regarding the technical limitations, we examine why a dramatic improvement in sensitivity cannot be achieved in the cases of $\varepsilon$ = $10^{-\frac{1}{2}}$ and $10^{-1}$, as shown in Figures \ref{fig:acc_10_to_negative_one_half} and \ref{fig:acc_10_to_negative_one}. In both of these cases, the arm cavity length that results in optimal sensitivity curve is shorter than DECIGO's default length of 1000\,km. The trade-offs of achieving better sensitivities by increasing the arm cavity lengths seems to be limited by higher vacuum fluctuations entering the system due to higher diffraction losses. As a result, a slightly shorter arm cavity length and a slightly higher cavity finesse, than the default design, yield the most optimal SNRs. An even shorter arm length is not optimal, as the influence of the acceleration noise gets more pronounced, limiting the total sensitivity that can be achieved in the presence of both diffraction losses and acceleration noises.

Finally, it is important to clarify the practical implications of these results for the DECIGO mission. While the SNR improvement remains modest under realistic acceleration noise levels ($\varepsilon \geq 10^{-1}$), this approach should be viewed as a complementary design strategy together with other quantum-noise reduction techniques, such as optical-spring quantum locking and squeezing, for detecting primordial gravitational waves in standard slow-roll inflation scenario. On the other hand, in more general scenarios where the constraints on tensor parameters are relaxed beyond those of standard slow-roll inflation, a wider parameter space for the primordial tensor spectrum remains viable. For example, analyses of Planck+BK15 data indicate that, when the inflationary consistency relation is not imposed, tensor tilts as large as $n_t \sim 2$ are still allowed within the 68\% confidence \cite{Planck2018inflation}. Such an extended tensor parameter space can naturally arise in models such as axion inflation \cite{Namba2016}, gauge field inflation \cite{Maleknejad2013}, and string gas cosmology \cite{Brandenberger2007}. In particular, string gas cosmology predicts a slightly blue-tilted primordial gravitational-wave spectrum. Therefore, although the sensitivity achieved in this study is insufficient for a definitive detection of the standard slow-roll inflationary background, the optimized DECIGO configuration may still provide non-trivial constraining power for more general models predicting enhanced gravitational-wave amplitudes in the deci-Hz frequency band.
\FloatBarrier
\suppressfloats[t]
\section{Conclusion}\label{sec:summary}

To improve the sensitivity of DECIGO, the use of optical springs and a homodyne detection scheme are considered. The optimal sensitivities are calculated in the presence of diffraction losses and acceleration noises. This is made possible by a recently developed framework that consistently incorporates diffraction-induced quantum state mixing in long-baseline cavities, which was not treated consistently in previous analyses. A rigorous treatment of quantum noise is adopted in the presence of diffraction losses in the arm cavities. Furthermore, the optimal design parameters are estimated using SNR towards primordial gravitational waves as a metric.

As a result, we have found that, in the presence of acceleration noise of level ($\varepsilon=10^{-\frac{1}{2}}$), which is part of DECIGO's default design - the sensitivity can be improved by approximately a factor of $1.5$ by employing an optical spring and a homodyne detection scheme. Furthermore, if $\varepsilon$ can be reduced to $10^{-2}$, an SNR of $33.4$ can be achieved, which is lower than the SNR of $79.6$ obtained in previous studies of an idealized quantum locking scheme with optical springs and homodyne detection under the assumption of negligible acceleration noise. Here, the SNR values are calculated assuming a primordial gravitational-wave energy density of ${\Omega}_{\mathrm{GW}} = 10^{-16}$, consistent with current CMB constraints. However, it should be noted that forming a very deep and sharp dip in the sensitivity curve is not suitable for achieving DECIGO's broader scientific objectives, and could potentially lead to reduced sensitivity due to other limiting technical noise sources, such thermal noise from residual gas.

Therefore, when these issues are comprehensively addressed, the optimized sensitivity corresponds to an SNR of about $8.18$ at $\varepsilon = 10^{-1}$. Although this represents an improvement over the configuration without detuned arm cavities, it remains insufficient to observe primordial gravitational waves. To ensure more reliable detection, it is therefore important to combine this approach with techniques that are not affected by diffraction losses, such as optical-spring quantum locking.

In conclusion, although this research demonstrates the potential for improving DECIGO's sensitivity toward the detection of primordial gravitational waves, the proposed approach is not sufficient to achieve a sensitivity, that can set a new upper limit. It is hoped that the results presented here will contribute to advancing observational capabilities in this field.

\blue{}

\vspace{0.5cm}
\noindent \textbf{Declaration of competing interest}\\[1mm]
The authors declare that they have no known competing financial interests or personal relationships that could have appeared to influence the work reported in this paper.

\vspace{0.5cm}
\noindent \textbf{Acknowledgments}\\[1mm]
We would like to thank Abhinav Patra for assistance for the English editing. This work was supported by JSPS KAKENHI, Grants Numbers JP22H01247/23K22518 and 25KJ1390.

\end{document}